# Universal Generative Modeling in Dual-domain for Dynamic MR Imaging

Chuanming Yu, Yu Guan, Ziwen Ke, Dong Liang, *Senior Member, IEEE*,
Qiegen Liu, *Senior Member, IEEE*

*Abstract*—Dynamic magnetic resonance image reconstruction from incomplete k-space data has generated great research interest due to its capability to reduce scan time. Nevertheless, the reconstruction problem is still challenging due to its ill-posed nature. Recently, diffusion models especially score-based generative models have exhibited great potential in algorithm robustness and usage flexibility. Moreover, the unified framework through the variance exploding stochastic differential equation (VE-SDE) is proposed to enable new sampling methods and further extend the capabilities of score-based generative models. Therefore, by taking advantage of the unified framework, we proposed a k-space and image Dual-Domain collaborative Universal Generative Model (DD-UGM) which combines the score-based prior with low-rank regularization penalty to reconstruct highly under-sampled measurements. More precisely, we extract prior components from both image and k-space domains via a universal generative model and adaptively handle these prior components for faster processing while maintaining good generation quality. Experimental comparisons demonstrated the noise reduction and detail preservation abilities of the proposed method. Much more than that, DD-UGM can reconstruct data of different frames by only training a single frame image, which reflects the flexibility of the proposed model.

## I. INTRODUCTION

Dynamic Magnetic Resonance Imaging (DMRI) is a key component of many clinical exams such as cardiac, perfusion, and functional imaging due to the characteristic of excellent spatial resolution and soft-tissue contrast [1]. However, the slow nature of the MR image acquisition scheme and the risk of peripheral nerve stimulation often restrict the achievable Spatio-temporal resolution and volume coverage in DMRI. Additionally, long scan durations make patients uncomfortable and increase the chance of producing motion artifacts. Therefore, it is of great necessity to accelerate DMRI and obtain high-quality images.

This work was supported in part by the National Natural Science Foundation of China under 61871206, 62122033. (C. Yu and Y. Guan are co-first authors.)
C. Yu, Y. Guan and Q. Liu are with Department of Electronic Information Engineering, Nanchang University, Nanchang 330031, China. ({yuchuanming, guanyu}@email.ncu.edu.cn, liuqiegen@ncu.edu.cn)
Z. Ke is with Institute for Medical Imaging Technology, School of Biomedical Engineering, Shanghai Jiao Tong University, Shanghai, 200030, China. (ziwen_ke@sjtu.edu.cn)
D. Liang is with Paul C. Lauterbur Research Center for Biomedical Imaging, SIAT, Chinese Academy of Sciences, Shenzhen 518055, China. (dong.liang@siat.ac.cn)

Over the past decades, constrained methods [2] have been proposed to accelerate MRI enforcing different prior information. Among them, compressive sensing (CS) [3-5] has represented a widely used accelerating approach, which uses a sparsity prior and incoherent sampling [6-11]. For example, k-t BLAST and k-t SENSE were developed to exploit signal correlations in k-t domain as prior information to recover non-acquired k-space data flexibly [4]. K-t FOCUSS proposed by Hong *et al.* took advantage of prediction and residual encoding, which estimated the residual signals from a very small number of k-t samples [8]. As an enhancement of k-t BLAST, [12] further exploited the relevant signal correlations as the temporal basis functions tailored to the training data, making the reconstruction problems inherently overdetermined. K-t ISD incorporated additional information on the support of the dynamic image in x-f space based on the theory of CS with partially known support [10].

In addition to utilizing the sparsity of prior information, many endeavors have been made to use low-rank as a prior regularization in DMRI reconstruction. It can complete missing or corrupted entries of a matrix by using low-rank and incoherence conditions [13-19]. A typical method on low-rank was put forward by Otazo *et al.* [11], where the nuclear norm was used to enforce low-rank in $L$, and the $L_1$ norm was used to enforce sparsity in $S$. The k-t SLR used the compact representation of the data in the Karhunen Louve transform domain to exploit signal correlations in k-t space and the sparsity of the data to improve the recovery rate further [15]. Some scholars also utilized dictionary learning as a way of excavating data sparsity. Liu *et al.* presented a novel sparse and dense hybrid representation (SDR) model which describes the sparse plus low-rank properties under dictionary learning [18]. Similarly, a novel method for the acceleration of cardiac DMRI was presented by Caballero *et al.* [19] which investigated the potential benefits of enforcing sparsity constraints on patch-based learned dictionaries and temporal gradients at the same time. These methods have made great progress in dynamic imaging and achieved improved results. However, iterative solution procedures require a relatively long time to achieve high-quality reconstructions, and most of these CS-based approaches do not utilize prior information from big data [20].

In comparison, deep learning (DL) methods are gaining popularity for their accuracy and efficiency. Unlike traditional approaches, the prior information and regularization are learned implicitly from data without specifying them in the training objective. DL methods can be roughly categorized into supervised [21-27] and unsupervised [28-32] schemes. So far, most DMRI reconstruction methods belong to the category of supervised learning [25-27]. For instance, Biswas *et al.* combined deep-learned priors along with complementary image regularization penalties which could reconstruct free-breathing and ungated cardiac MRI data from highly under-sampled multi-channel measurements [27]. However, the supervised algorithms require a large training dataset with accurate labels to learn a network, which is difficult in some special circumstances. Given the defect, unsupervised DL methods were developed which learn the distribution of data only from the measured data. For example, Nguyen-Duc *et al.* [33] recovered high-frequency information using a shared 3D convolution-based dictionary built progressively during the reconstruction process, while low-frequency information was recovered using a total variation-based energy minimization method that leverages temporal coherence in DMRI.

Recently, score-based models [34-37] have gained wide interest as a new class of generative models. This model achieved surprisingly high sample quality without adversarial training [38-40]. Among many works, score matching [34] attempted to match the derivative of the model density with the derivative of the data density. This approach has recently been successfully applied to medical image reconstruction [41]. Subsequently, Song *et al.* studied a generalized discrete score-matching procedure with a continuous stochastic differential equation (SDE), subsuming diffusion models into the same framework.

Despite all the successes achieved by the aforementioned methods, it is easy to discover that they principally work in the image domain while only a handful of approaches can be employed in dual-domain. The applicability of deep learning models to dual-domain is yet to be fully explored. In this work, we propose a **D**ual-**D**omain **U**niversal **G**enerative **M**odel (DD-UGM) which exploits the potentials of k-space and image dual-domains jointly. Firstly, the algorithm takes advantage of the noise distribution of the SDE framework as a prior regularization. Subsequently, the operation of Predictor-Corrector (PC) in the generative model is used as a numerical SDE solver to generate samples. Moreover, singular value decomposition (SVD) is also employed in the k-space domain as the second prior regularization to exploit the sparsity in both spatial and temporal domains. In general, DD-UGM combines the score-based generative model with other traditional methods as a novel constrained optimization item in the reconstruction stage and then solves it through an alternating minimization scheme.

The main contributions of this work are summarized as follows:
- *Generation via different regularizations:* A novel model DD-UGM which takes advantage of the prior of the dual-domain obtained by score-based generative framework is proposed to reconstruct images. Furthermore, another low-rank constraint along the time dimension in k-space is introduced to improve the visual quality of reconstruction results.
- *Generative learning in dual-domain*: A hybrid-domain method that can process the complementary prior knowledge generated from the k-space and image domains is presented for high-precision DMRI reconstruction.
- *Universal and flexible generative modeling*: Since the low-rank constraint adequately utilizes the redundant information along the time dimension of the dynamic MR images, DD-UGM can be used to reconstruct images of any number of frames just by training the single frame image. It indicates the generalizability and potential task flexibility of the algorithm.

The rest of the thesis is organized as follows. Section II illustrates the background knowledge, including DMRI. Section III introduces the DD-UGM and details of the reconstruction process. We demonstrate the performance of our reconstruction method and compare it with classic and deep learning methods in Section IV. Finally, discussions and conclusions are given in Sections V and VI, respectively.

## II. PRELIMINARIES

### A. DMRI Acquisition

Considering a Cartesian k-space trajectory where $k_x$ denotes the phase encoding direction, $k_y$ denotes the readout direction. $u(x,t)$ denotes the image domain content at $x$ and time $t$. The k-space measurement $v(k,t)$ is then formulated as:

$$v(k,t) = \int \sigma(x,t)e^{-j2\pi kx}dx = \iint \rho(x,f)e^{-j2\pi(kx+ft)}dxdf \quad (1)$$

where $\rho(x,f)$ is the 2D spectral signal in x-f domain. This can also be represented in a matrix form: $\mathbf{v} = F\rho$, in which $\mathbf{v}$ and $\rho$ stand for the stacked k-t space measurement vectors and x-f image, respectively. $F$ performs a 2D discrete Fourier transform along the x-f direction.

Alternatively, $u(x,t)$ denotes the spatio-temporal signal, the DMRI measurements correspond to the samples of the signal in k-t space, corrupted by noise $n$. Hence, the above under-sampled measurement can be modeled in a vector form as:

$$a(x,t) = F_u u(x,t) + n; \quad b(k,t) = M_u v(k,t) + n \quad (2)$$

where $n \in \mathbb{C}^{M \times N}$ is additive white Gaussian acquisition noise. $M$ is the total number of voxels in the image and $N$ is the number of image frames. The DMRI measurements correspond to the samples of the signal in k-t space is under-sampled, which is a subset $\Omega$ of k-space. $M_u$ denotes the under-sampled operator to acquire only a subset of k-space, which contains the rows from the identity matrix that corresponds to the samples that are in $\Omega$. $F_u$ is an under-sampling Fourier encoding matrix.

### B. Reconstruction with Low-rank Prior

Since the problem of DMRI reconstruction is ill-posed and requires a regularization term to assist the solution, many CS-based methods are proposed to exploit the temporal correlation in DMRI reconstruction. Due to the k-t space MR signal is modeled as being partially separable along the spatial and temporal dimensions, which results in a rank-deficient data matrix. Obviously, image reconstruction is then formulated as a low-rank matrix recovery problem, which can be solved via emerging low-rank matrix recovery techniques. Therefore, in order to reconstruct $v(k,t)$ of Eq. (2), the ill-posed problem is constrained by adding some low-rank prior knowledge and solving the following optimization problem:

$$v^* = \arg\min_v \left\{ \left\| M_u(v(k,t)) - b(k,t) \right\|_2^2 + \beta \left\| v(k,t) \right\|_* \right\} \quad (3)$$

The first term is data consistency which ensures that the k-space of reconstruction is consistent with the actual measurements in k-space. The second term is often referred to as the prior regularization. $\beta$ is the regularization parameter. In the methods of CS, $\left\| \cdot \right\|_*$ is usually a low-rank constraint of $v(k,t)$ in some transform domains.

### C. *Probabilistic Generative Model*

Score-based generative models have recently contributed greatly to reconstructing medical images and could be flexibly adapted to different measurements during test time. Score-based approaches define a forward diffusion process for transforming data into noise and generating data by reversing it. Song *et al.* [38] designed a brand-new model, which leads to record-breaking performance for unconditional image generation, called the score-based generative model through SDE. Its framework encapsulates previous approaches in score-based generative modeling and diffusion probabilistic modeling, allowing for new sampling procedures and new modeling capabilities.

Two successful classes of probabilistic generative models involve sequentially corrupting training data with slowly increasing noise, and then learning to reverse this corruption in order to form a generative model of the data. In particular, the first category is score matching with Langevin dynamics estimates the score, *i.e.*, the gradient of the log probability density with respect to data at each noise scale, and then uses Langevin dynamics to sample from a sequence of decreasing noise scales during generation. Denoising diffusion probabilistic modeling (DDPM) of the second category trains a sequence of probabilistic models to reverse each step of the noise corruption, using knowledge of the functional form of the reverse distributions to make training tractable. Moreover, the DDPM training objective implicitly computes scores at each noise scale for continuous state spaces. We therefore refer to these two model classes together as score-based generative models.

### D. *Learning in Dual-domain Strategy*

Dual-domain model has been widely used in the field of MRI reconstruction which can complementarily leverage prior information presented in k-space and image domains. A previous study proposed a hybrid model, which was trained end-to-end in a supervised manner, and was assessed only on single-coil data [42]. Utilizing the generative model to prior learning, wang *et al.* introduced a dual-domain generative adversarial network, which added the regularization in the frequency domain to correct the high-frequency imperfection [43]. Recently, by using energy function as the generative model, Tu *et al.* proposed a k-space and image domains collaborative generative model termed KI-EBM to comprehensively estimate the MR data from under-sampled measurement [31]. Particularly, carrying out the combination modes of k-space and image dual-domain in parallel and sequential orders was explored. Based on the success of the above methods, in this work we propose a hybrid approach that works with information captured from k-space and image dual-domain. The main difference is that we adopt the score-based model for prior learning, rather than another generative model.

## III. PROPOSED METHOD

We first introduce the theory of universal generative modeling in k-space and image dual-domain (Sec. A). Note that the weighting strategy for tackling the issue in k-space domain is also presented. Then, the data distribution of the score-based network is introduced as a prior regularization for better image generation (Sec. B). Specially, a traditional Hankel low-rank algorithm which restricts the information of the temporal dimension of k-space data is utilized to increase the generative ability of k-space domain. Finally, with regard to the resulting mathematical model, an alternating minimization scheme is adopted to address it.

### A. *Prior Learning in Dual-domain*

Generally, the core idea of this study is to use the universal generative model to capture certain statistical properties of the input data. Due to the input data of different domains contains different information, we employ a framework DD-UGM for score-based generative modeling based on VE-SDE to learn the prior information separately for the k-space and image dual-domain. Such a model does not only perform on one domain but on dual-domain. As a result, the DD-UGM can leverage the information of dual-domain to increase the details of generated images further.

***Learning of Prior Distribution in K-space Domain:*** In order to make the generated model better extract the prior information in k-space domain, we artificially suppress low-frequency prior information through a weighted matrix $\mathbb{W}$. It is worth noting that the weighted technology makes the output of high-frequency information sparser. This property is consistent with the spirit of CS-MRI. Therefore, we apply k-space weighting technology to input data to construct more effective and robust prior information, which can be represented as:

$$v_{\mathbb{W}}(k,t) = v(k,t) \odot \mathbb{W}; \quad v(k,t) = v_{\mathbb{W}}(k,t) / \mathbb{W} \quad (4)$$

where $v(k,t)$ denotes k-space data and $\mathbb{W} = (p \times (R_x^2 + R_y^2))^q$. $R_x$ and $R_y$ are the count of frequency encoding lines and phase encoding lines, respectively. $p$ and $q$ stand for two parameters to adjust the weight, respectively. For ease of presentation, we hereafter denote by $v$ the $v(k,t)$, and use the $v_{\mathbb{W}}$ to indicate the $v_{\mathbb{W}}(k,t)$.

As mentioned above, perturbing data with multiple noise scales is key to the success of previous generative methods. Therefore, it is very necessary to generalize this idea further to an infinite number of noise scales, such that perturbed data distributions evolve according to an SDE as the noise intensifies. Our goal is to construct a diffusion process $\{v_{\mathbb{W}}(t)\}_{t=0}^T$ indexed by a continuous-time variable $t \in [0,T]$, such that $v_{\mathbb{W}}(0) \sim p_0$, for which we have a dataset of samples, and $v_{\mathbb{W}}(T) \sim p_T$, for which we have a tractable form to generate samples efficiently. In other words, $p_0$ is the data distribution and $p_T$ is the prior distribution. This diffusion process can be modeled as the solution to the following SDE:

$$dv_{\mathbb{W}} = f(v_{\mathbb{W}}, t)dt + g(t)dw \quad (5)$$

where $w$ is the standard Wiener process (*a.k.a.*, Brownian motion), $f(\cdot, t): \mathbb{R}^d \to \mathbb{R}^d$ is a vector-valued function called the drift coefficient of $v_{\mathbb{W}}(t)$, and $g(\cdot): \mathbb{R}^d \to \mathbb{R}^d$ is a scalar function known as the diffusion coefficient of $v_{\mathbb{W}}(t)$. It is notably that the SDE has a unique strong solution as long as the coefficients are globally Lipschitz in both state and time [44]. We hereafter denote by $p_t(v_{\mathbb{W}})$ the probability density of $v_{\mathbb{W}}(t)$, and use $p_{st}(v_{\mathbb{W}})$ to denote the transition kernel from $v_\mathbb{W}(s)$ to $v_\mathbb{W}(t)$, where $0 \leq s < t \leq T$.

Typically, there are various ways of designing the SDE in Eq. (5) by choosing different functions for $f(\cdot)$ and $g(\cdot)$, such that it diffuses the data distribution into a fixed prior distribution. According to the work of Song et al. [28], we use the Variance Exploding SDE (VE-SDE) by choosing $f = 0$, $g = \sqrt{d[\sigma^2(t)]/dt}$ to form the following manner:

$$dv_\mathbb{W} = \sqrt{\frac{d\left[\sigma^2(t)\right]}{dt}} dw \tag{6}$$

In the limit of $N \to \infty$, $\{\sigma_i\}_{i=1}^N$ becomes a Gaussian noise function $\sigma(t)$ with a variable in continuous time $t \in [0,1]$, which can be redescribed as a positive noise scale $\{\sigma_i\}_{i=1}^N$. The key idea of VE-SED in k-space domain is visualized in Fig. 1.

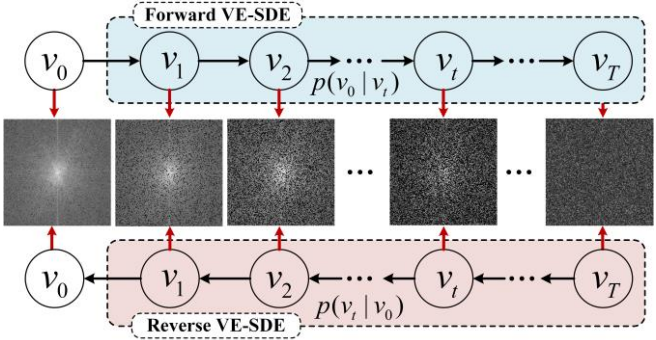

**Fig. 1.** VE-SDE smoothly transforms a data distribution to a known prior distribution by slowly injecting noise in the k-space domain, and a corresponding reverse-time VE-SDE that transforms the prior distribution back into the k-space data distribution by slowly removing the noise.

Afterward, the score of data distribution in k-space domain $\nabla_{v_\mathbb{W}} \log p_t(v_\mathbb{W})$ can be estimated by training the score-based generative model based on VE-SDE. The parameters optimization process of the model $s_\theta(v_\mathbb{W}(t), t)$ is as follows:

$$\theta^*_{v_\mathbb{W}} = \arg\min_{\theta_{v_\mathbb{W}}} \mathbb{E}_t \left\{ \lambda_{v_\mathbb{W}}(t) \mathbb{E}_{v_\mathbb{W}(0)} \mathbb{E}_{v_\mathbb{W}(t)|v_\mathbb{W}(0)} \right. \tag{7}$$
$$\left. [\left\| S_{\theta_{v_\mathbb{W}}}(v_\mathbb{W}(t),t) - \nabla_{v_\mathbb{W}(t)} \log p_{0t}(v_\mathbb{W}(t)|v_\mathbb{W}(0))\right\|_2^2] \right\}$$

where $\lambda_{v_\mathbb{W}}(t)$ is a positive weighting function, $v_\mathbb{W}(0) \sim p_0(v_\mathbb{W})$ and $v_\mathbb{W}(t) \sim p_{0t}(v_\mathbb{W}(t)|v_\mathbb{W}(0))$. Since VE-SDE has affine drift coefficients, its perturbation kernels $p_{0t}(v_\mathbb{W}(t)|v_\mathbb{W}(0))$ are all Gaussian and can be computed in closed-forms. It makes training process efficient.

***Learning of Prior Distribution in Image Domain:*** As we all know, the amplitude values under two neighbor pixels are very different and differ each other very significantly in k-space domain. Thence, a weighted operation on the input k-space data is utilized for the training of the score-based generative model. By contrast, the amplitude values under different pixel location in image domain are homogeneous and the range of them is very close. To this extent, we directly obtain the prior information of the dynamic MR images in the image domain. Specifically, instead of training the model in k-space domain, we also consider an image domain content $u(x,t)$ as a single target for the denoising process instantly.

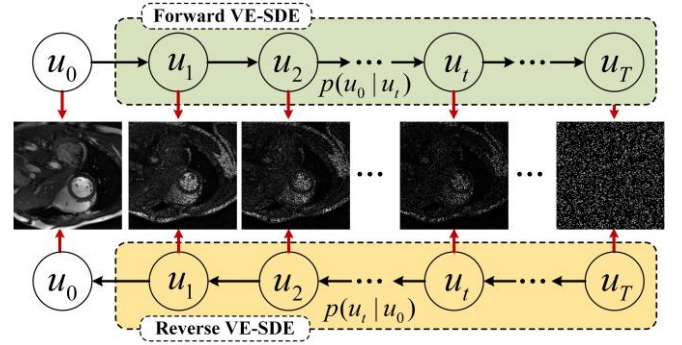

**Fig. 2.** The corresponding bidirectional process of VE-SDE in image domain, which performs a slow noise injection process and noise removal process on dynamic MR images.

Subsequently, by assembling the discussions in the above subsections, the diffusion process can be modeled in image domain as the solution to the following VE-SDE:

$$du = \sqrt{\frac{d\left[\sigma^2(t)\right]}{dt}} dw \tag{8}$$

The corresponding bidirectional process of VE-SDE in image domain is described in Fig. 2. Obviously, to estimate the score of a distribution in image domain $\nabla_u \log p_t(u)$, we can train another score-based model $s_{\theta_u}(u(t),t)$ on image samples with VE-SDE as follows:

$$\theta^*_u = \arg\min_{\theta_u} \mathbb{E}_t \left\{ \lambda_u(t) \mathbb{E}_{u(0)} \mathbb{E}_{u(t)|u(0)} \right. \tag{9}$$
$$\left. [\left\| S_{\theta_u}(u(t),t) - \nabla_{u(t)} \log p_{0t}(u(t)|u(0))\right\|_2^2] \right\}$$

Similar to the notation in previous content, $\lambda_u(t)$ is a positive weighting function, $t$ is uniformly sampled over $[0,T]$, with sufficient data and model capacity, VE-SDE ensures that the optimal solution to Eq. (9). In summary, we optimize the score-based generative model through the prior information of the dual-domain. To a certain extent, two different models are obtained by prior learning which allows reconstruction information to be shared across the multiple iterations of the process.

### B. DMRI Reconstruction in Dual-domain

As aforementioned, the proposed DD-UGM is formed of two inner-models that learn complementary features from k-space and image domains separately to generate samples. Thus, in order to exploit the dynamic nature and temporal redundancy of different data in the k-space domain and image domain, we further propose to jointly model in the dual-domain for dynamic MRI reconstruction. Considering the unique nature of k-space data, different patterns are excavated in the reconstruction phase. These components are described in the following subsections.

***K-space Prior for DMRI Reconstruction:*** Regularization related to the data prior is a key component in the reconstruction problem to reduce imaging artifacts and improve imaging precision. Specifically, the reconstruction task is normally formulated as solving an optimization problem with two terms, *i.e.*, data fidelity and regularization. Notably, two different regularization terms are combined to promote better image generation in the reconstruction process of k-space domain. The deep generation prior of the k-space data, obtained through a VE-SDE model, is inserted into the reverse VE-SDE as the first regularization term. Additionally, the traditional low-rank algorithm by truncating the SVD representation is introduced in the process of regularization.

**Step 1: Prior regularization.** After training a time dependent score-based model $S_\theta(v_\mathbb{W}(t),t)$, we can use it to construct the reverse-time VE-SDE and then simulate it with numerical approaches to generate samples from $p_0$, given by:

$$dv_\mathbb{W} = -d\sigma^2(t) \cdot \nabla_{v_\mathbb{W}} \log p_t(v_\mathbb{W}) + \sqrt{d[\sigma^2(t)]/dt}\, d\overline{w} \quad (10)$$

where $\overline{w}$ is a standard Wiener process in reverse-time direction, and $dt$ is an infinitesimal negative time step.

By reversing VE-SDE, we can convert random noise into data for sampling. For sample update step, the predictor-corrector (PC) sampling is used in this work. The predictor (P) refers to a numerical solver for the reverse-time SDE. Specifically, the samples from the prior distribution in k-space domain can be obtained from the reverse VE-SDE in Eq. (10), which can be discretized as follows:

$$v_{\mathbb{W}_i} = v_{\mathbb{W}_{i+1}} + (\sigma_{i+1}^2 - \sigma_i^2) s_\theta(v_{\mathbb{W}_{i+1}}, \sigma_{i+1}) + \sqrt{\sigma_{i+1}^2 - \sigma_i^2}\, z$$
$$i = I-1, \cdots, 0 \quad (11)$$

where $z \sim N(0,1)$, $v_\mathbb{W}(0) \sim p_0$, and we set $\sigma_0 = 0$ to simplify the notation. Corrector (C) is used after the predictor to obtain a more efficient and robust iterative formulation. Langevin dynamics is considered as the C, which transforms any initial sample $v_\mathbb{W}(0)$ to an approximate sample $p_t(v_\mathbb{W})$ with the following procedure:

$$v_{\mathbb{W}_{i,j}} = v_{\mathbb{W}_{i,j-1}} + \varepsilon_i s_\theta(v_{\mathbb{W}_{i,j-1}}; \sigma_i) + \sqrt{2\varepsilon_i}\, z;\quad j=1,2,\cdots,J \quad (12)$$

where $\varepsilon_i > 0$ is the step size, and $z \sim N(0,1)$ is standard normal distribution. The theory of Langevin dynamics guarantees that when $j \to \infty$ and $\varepsilon_i \to 0$, $v_{\mathbb{W}_{i,j}}$ is a sample from $p_t(v_\mathbb{W})$ under some regularity conditions. After PC operation, we obtain the k-space data generated by the network. It is worth noting that the output of the universal generative model needs to be divided by the weight matrix $\mathbb{W}$, and then we construct another regularization on it.

**Step 2: Structure regularization.** To reconstruct $v(k,t)$ and obtain high-quality dynamic reconstruction images, we construct the data matrix from the prior information generated in the k-space domain and analyze the hard-threshold singular values of the data matrix. Once the dynamic k-space data matrix is transformed to Hankel matrix, we can apply SVD to it to decompose the information into signal and noise subspaces. Subsequently, hard-threshold value is processed. A low-rank constraint is applied to solve the following optimization problem:

$$\min_k \| Mv(k,t) - b(k,t) \|_2^2$$
$$s.t.\ rank(A) = a,\ v = H^+(A) \quad (13)$$

where $H^+$ is pseudo-inverse operator, $A$ is a data matrix enforced low-rank by setting hard-threshold singular values and $a$ is the rank of the data matrix. At that time, the spectral penalty is non-convex, which can use fewer measurement data to improve the reconstruction quality.

*Image Prior for DMRI Reconstruction:* A few adjustments are made in processing the regularization for better image domain reconstruction. During the reconstruction, it is noticed that due to the special nature of the image data, the structural regularization term is not used to restrict it quadratically. In detail, it is the reconstruction process with only one regularization in the image domain, that is, the prior of the image obtained by the universal generative model is directly embedded into the reconstruction process for iterative generation. The pipelines of the reconstruction process in the k-space domain and the image domain are illustrated in Fig. 3.

Similarly, after the time-dependent score-based model $S_\theta(u(t),t)$ is trained in the image domain, we can represent the reverse time VE-SDE as:

$$du = -d\sigma^2(t) \cdot \nabla_u \log p_t(u) + \sqrt{d[\sigma^2(t)]/dt}\, d\overline{w} \quad (14)$$

Accordingly, the PC sampler is introduced to correct errors for the evolution of the discretized reverse-time SDE in image domain. For concision of presentation, we directly give the process of PC sampling from the perspective of the mathematical as follows:

$$\begin{cases} P:\ u_i = u_{i+1} + (\sigma_{i+1}^2 - \sigma_i^2) s_\theta(u_{i+1}, \sigma_{i+1}) + \sqrt{\sigma_{i+1}^2 - \sigma_i^2}\, z \\ \qquad i = I-1,\cdots,0 \\ C:\ u_{i,j} = u_{i,j-1} + \varepsilon_i s_\theta(u_{i,j-1}; \sigma_i) + \sqrt{2\varepsilon_i}\, z \\ \qquad j = 1,2,\cdots,J \end{cases} \quad (15)$$

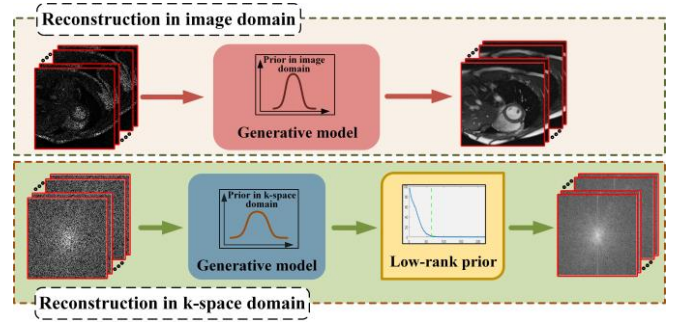

**Fig. 3.** The pipeline of the reconstruction process in the k-space domain and image domain. Note that the generative model is trained with different prior information.

*Data Consistency (DC):* For the purpose of utilizing the strong generation ability of the generated model and generating fixed data, a data consistency module is added after the regularization. Generally, data consistency is simultaneously enforced in each iterative reconstruction step to ensure that the output is consistent with the original k-space information. Due to the dual-domain learning strategy that optimizes $v^*$ and $u^*$ recurrently by different generative models employed in this work, the output of the different models needs to be weighted before data consistency, which facilitates the subsequent hybrid prior information to be fed into the network again. Denoting the virtual variable as $g^*$, the final optimization target can be written as:

$$g^* = \eta_1(v^*) + \eta_2 \mathcal{F}(u^*) \quad (16)$$

where $\eta_{(\cdot)}$ controls the level of linear combination between k-space domain values and image domain values. $\mathcal{F}$ is the Fourier transform. As a result, the corresponding output from data consistency layer can be thus formulated as:

$$g(k,t) = \begin{cases} g(k,t) & \text{if } k \notin \Omega \\ \dfrac{g(k,t) + \mu g_u(k,t)}{1+\mu} & \text{if } k \in \Omega \end{cases} \quad (17)$$

where $\Omega$ denotes an index set of the acquired k-space samples. $g(k,t)$ represents an entry at index $j$ in k-space generated by network and $g_u(k,t)$ is the un-der-sampled k-space measurement. In the noiseless setting (*i.e.*, $\mu \rightarrow \infty$), we replace the $k$-th predicted coefficient with the original coefficient if it has been sampled.

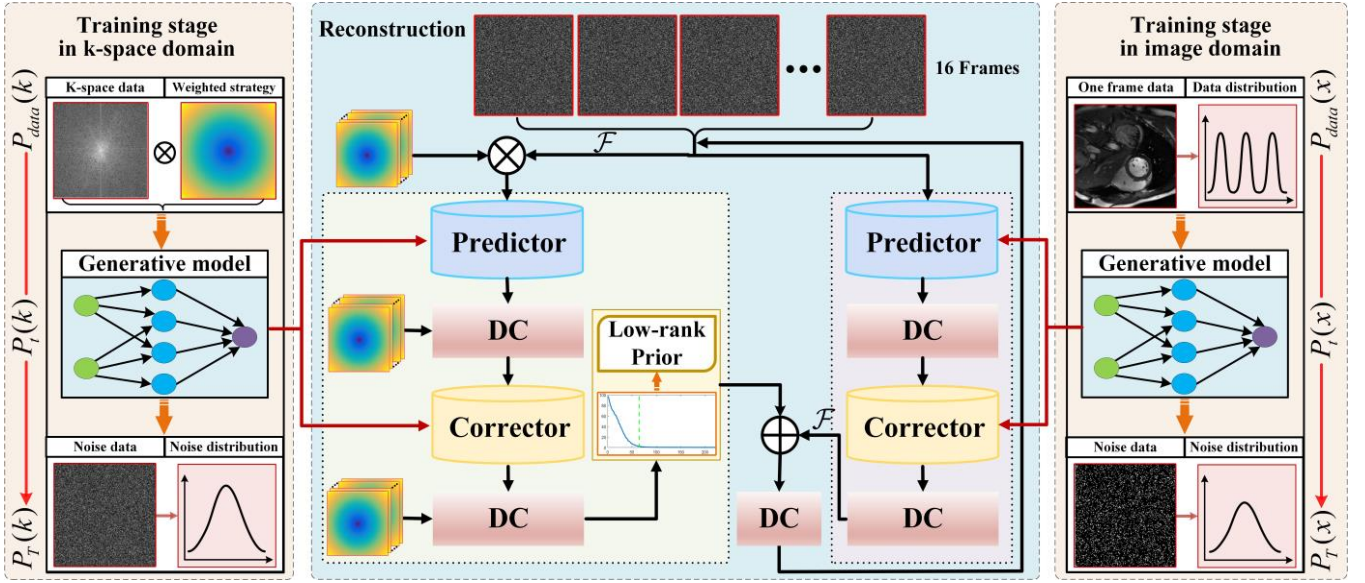

**Fig. 4.** The proposed DD-UGM method in dual-domain for DMRI reconstruction. Left: Universal generative model to learn the k-space prior information via denoising score matching. Middle: Reconstruction to progressively remove aliasing and recover fine details via PC operation and low-rank prior. Right: Universal generative model to learn the image prior information via denoising score matching.

### C. Summary of DD-UGM

In summary, the prior information of **Dual-domain** is mined by a novel **U**niversal **G**enerative **M**odel for DMRI reconstruction. The overall pipeline of our algorithm DD-UGM is illustrated in Fig. 4. It consists of two major parts, *i.e.*, prior learning and reconstruction stages in k-space and image dual-domain. Our intention is to learn the prior distribution via score-based generative model in the training stage. More precisely, due to the unique nature of the k-space domain, we exploit a weighted matrix multiplied by the under-sampled single frame k-space data as the network input, while using only the single frame image as the input to the image domain model. It means that we will obtain two models with different prior information. Interestingly, as the number of iterations increases, the parameters of the two models are updated by the hybrid information, which indicates that we jointly exploit the complementary properties of the dual-domain prior. In the reconstruction phase, the under-sampled k-space data from all frames are used for testing, and the outputs are the reconstructed dynamic images. Moreover, the aliasing of under-sampled cardiac MR images is removed progressively via iteratively applying predictor and corrector steps. Note that we also implement the SVD algorithm on Hankel transformed multiple frames data as an additional constraint to recover fine image details in k-space domain. To ensure that the k-space of the reconstructed data is consistent with the actual measurements, data consistency is eventually imposed. Furthermore, **Algorithm 1** explains the reconstruction algorithm in detail.

| Algorithm 1: DD-UGM |
|---|
| **Training stage** |
| **1. Input:** k-space and image datasets: $v(k,t)$, $u(x,t)$ |
| **2. Construct weighted data matrix:** $v_{\mathbb{W}}(k,t) = v(k,t) \odot \mathbb{W}$ |
| **3. Training:** Eq. (7) and Eq. (9) |
| **4. Output:** $S_{\theta_{v_{\mathbb{W}}}}(v_{\mathbb{W}}(t),t)$, $S_{\theta_u}(u(t),t)$ |

| Reconstruction stage |
|---|
| **Setting:** $I, J, \mathbb{W}, v, u, \sigma, \varepsilon$ |
| 1: $v^I \sim \mathbb{N}(0, \sigma_T^2 \mathcal{I})$, $u^I \sim \mathbb{N}(0, \sigma_T^2 \mathcal{I})$ |
| 2: **for** $i = I-1$ to $0$ **do (Outer loop)** |
| 3: $\quad v_{\mathbb{W}}^i \leftarrow$ Predictor $(v_{\mathbb{W}}^{i+1}, \sigma_i, \sigma_{i+1}); u^i \leftarrow$ Predictor $(u^{i+1}, \sigma_i, \sigma_{i+1})$ |
| 4: $\quad v^i \leftarrow v_{\mathbb{W}}^i / \mathbb{W}$ |
| 5: $\quad$ Update $v^i, u^i$ via Eq. (17) |
| 6: $\quad v_{\mathbb{W}}^i \leftarrow v^i \odot \mathbb{W}$ |
| 7: $\quad$ **for** $j = 1$ to $J$ **do (Inner loop)** |
| 8: $\quad\quad v_{\mathbb{W}}^{i,j} \leftarrow$ Corrector $(v_{\mathbb{W}}^{i,j-1}, \sigma_i, \varepsilon_i); u^{i,j} \leftarrow$ Corrector $(u^{i,j-1}, \sigma_i, \varepsilon_i)$ |
| 9: $\quad$ **end** |
| 10: $\quad v^i \leftarrow v_{\mathbb{W}}^i / \mathbb{W}$ |
| 11: $\quad$ Update $v^i, u^i$ via Eq. (17) and obtain $v^*, u^*$ |
| 12: $\quad A_v^* \leftarrow H(v^*)$ (Hankel matrix) |
| 13: $\quad [U, S, V] = svd(A_v^*)$ (Perform SVD) |
| 14: $\quad A_v^* \leftarrow US_r V^T$ (Low-rank matrix) |
| 15: $\quad v^* \leftarrow H^+(A_v^*)$ (Transform matrix back to k-space data) |
| 16: $\quad g^* = \eta_1(v^*) + \eta_2 \mathcal{F}(u^*)$ (Weighted in dual-domain) |
| 17: $\quad$ Update $g^*$ via Eq. (17) (Data consistency) |
| 18: **end** |
| 19: **Return** $g^{rec} \leftarrow g^*$ |

## IV. EXPERIMENTS

In this section, the performance of the proposed method is demonstrated in a variety of image styles, compared with different state-of-the-art algorithms on different acceleration rates. The comparative experiments validate the effective reconstruction ability of the proposed algorithm. Furthermore, we have unified the methods of calculating performance metrics for a fair comparison. The source code of the DD-UGM is available at: https://github.com/yqx7150/DD-UGM.

## A. Data Acquisition

The fully-sampled cardiac MR dataset was obtained on a scanner (MAGNETOM Trio, Siemens Healthcare, Erlangen, Germany) with 20 coil elements total from 30 healthy volunteers. For each subject, 10 to 13 short-axis slices were imaged with the retrospective electrocardiogram (ECG)-gated segmented bSSFP sequence during breath-holding. The protocol was approved by the Institutional Review Board (IRB) of Shenzhen Institutes of Advanced Technology, and informed consent was obtained from each volunteer. The following sequence parameters: $FOV = 330 \times 330$ mm, acquisition matrix $= 256 \times 256$, slice thickness $= 6$ mm, TR/TE $= 3.0$ ms/$1.5$ ms. The temporal resolution was $40.0$ ms, and each data point had approximately 25 phases that covered the entire cardiac cycle. The scan parameters were kept the same across all patients.

The single-coil data was used in the experiment. Specifically, utilizing the adaptive coil combining method, the raw multi-coil data of each frame were combined to produce single-coil complex-valued images [45]. We cropped the images along the $x$ and $y$ directions with size of $192 \times 192$ for training. Additionally, the time-series data consists of 16 frames uniformly. Finally, we only used 200 single-coil 2D-t cardiac MR data for training and 10 data for testing, and each data contains 16 frames.

In addition to retrospective reconstruction, we also tested our network on a real-time prospectively acquired OCMR dataset [46]. The prospective cardiac cine dataset contains two parts: 57 slices of fully-sampled training data for fine-tuning and 7 prospectively under-sampled data at 9-fold acceleration for the test. Both the fully-sampled data and under-sampled data were collected in the short-axis and long-axis view on a 3T Siemens MAGNETOM Prisma machine with a 34-channel receiver coil array, using a bSSFP sequence with $FOV = 800 \times 300$ mm, acquisition matrix is $384 \times 144$, slice thickness is $8$ mm, TR is $38.4$ ms and TE is $1.05$ ms. The under-sampled data was collected in a real-time mode under free-breathing conditions. Because of the difference in the scan scheme, the fully-sampled data had an average frame of 18, and the under-sampled data had 65 frames for each. More details can be found in [46].

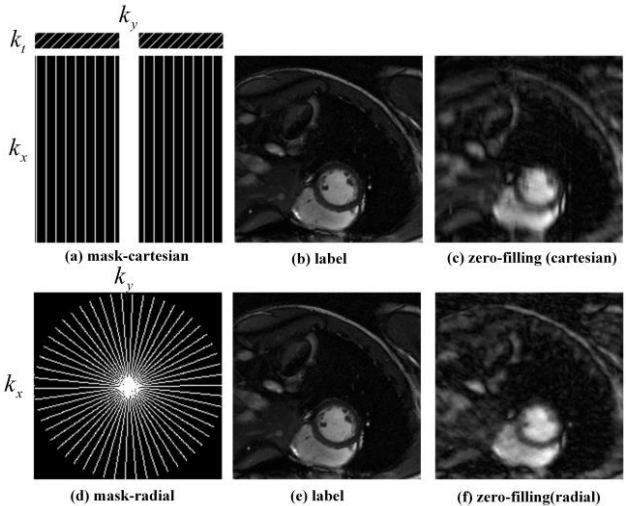

**Fig. 5.** Visualization of some experimental data and representative sampling masks.

Fig. 5 illustrates two sampling trajectories of the training and testing data. The Cartesian sampling pattern with time-interleaved is chosen to obtain the specified acceleration factor. Moreover, we consider Radial sampling pattern with uniform angular spacing for each frame of cardiac cine sequences. It not only achieves greater aliasing incoherence, but also ensures that there are not any sudden jumps across the samples acquired over time.

## B. Model Configuration and Implementation

The framework of DD-UGM consists of two different models, which are trained by inputting weighted k-space data and original dynamic MR images respectively. The output of the network is the perturbed noise distribution in k-space and image dual-domain. Accordingly, the hyper-parameters in the network are described as follows: The network parameters of k-space and image domains are the same during training. Two models (image domain model and k-space domain model) are trained with an exponentially decaying learning rate which initial value is set to $0.005$. In addition, the Adam optimizer with parameter $\beta_1 = 0.9$ and $\beta_2 = 0.999$ is used in this experiment. For noise variance schedule, we fix $\sigma_{max} = 378$, $\sigma_{min} = 0.01$ and $r = 0.075$ in the training phase while $\sigma_{max} = 4$, $\sigma_{min} = 0.01$ and $r = 0.075$ are chosen in the reconstruction procedure. As for the weighting operation, the k-space and image domains weight ratio are set as $\eta_1 = 0.75$ and $\eta_2 = 0.25$. Moreover, the iteration step size is set $I = 1000$, $J = 1$ respectively.

The involved parameters in the experiment are set following the guidelines in comparison method's original papers. Additionally, the training and testing phase of DD-UGM model is implemented on an Ubuntu 14.04 LTS (64-bit) operating system equipped with an Intel Core i7-4790K central processing unit (CPU) and NVIDIA GTX1080 graphics processing unit (GPU, 8 GB memory) in the open framework PyTorch with CUDA and CUDNN support.

## C. Performance Metrics

To evaluate the performance of the proposed network, we compare the reference images with the reconstructed images using both qualitative and quantitative metrics. For qualitative metrics, we use visual inspection and emphasize the image regions that are better or worse. Specifically, the peak-signal-to-noise ratio (PSNR), structural similarity index (SSIM), and mean squared error (MSE) are employed as our quantitative metrics. $x_{ori}$ and $x_{rec}$ denote the ground truth and reconstructed image, respectively.

The PSNR describes the relationship of the maximum possible power of a signal with the power of noise corruption, which is commonly used as a form of fidelity measure. Specifically, PSNR is expressed as:

$$PSNR = 20\log_{10}[\text{Max}(x_{ori})/\|x_{ori} - x_{rec}\|_2] \quad (18)$$

The SSIM is a perceptual metric introduced in [47], which indicates visual perception and the ability of detail preservation in the reconstructed image. It can be defined as:

$$SSIM = \frac{(2\mu_{x_{ori}}\mu_{x_{rec}} + c_1)(2\sigma_{x_{ori}x_{rec}} + c_2)}{(\mu_{x_{ori}}^2 + \mu_{x_{rec}}^2 + c_1)(\sigma_{x_{ori}}^2 + \sigma_{x_{rec}}^2 + c_2)} \quad (19)$$

The MSE is chosen to evaluate the overall accuracy of the reconstruction quality, which is measured as:

$$MSE = \|x_{ori} - x_{rec}\|_2^2 \quad (20)$$

### D. DMRI Reconstruction Results

In this subsection, we describe the quantitative and qualitative reconstruction comparisons of the proposed algorithm with 4 representative DMRI methods, including conventional low-rank constraint methods: BCS [1], DLTG [19], k-t SLR [15], and deep-learning-based approach 3D-CSC [33]. Moreover, to further demonstrate the reconstruction performance of the proposed method, the comparative experiment is conducted at different acceleration rates (8-fold and 10-fold). For data processing, we focus on the Cartesian sampling pattern with time-interleaved, which is chosen to obtain the specified acceleration factor. It not only achieves greater aliasing incoherence, but also ensures that there are not any sudden jumps across the samples acquired over time.

The quantitative evaluation results of reconstruction methods under different acceleration factors are provided in Table I. All the quantitative results are the averaged results on the 16 frames test datasets. As seen in Table I, it can be observed that the PSNR, SSIM, and MSE indexes of the DD-UGM perform better than BCS, k-t SLR, and 3D-CSC methods in the case of 8-fold and 10-fold sampling rates. Specifically, the DD-UGM model can offer around 2 dB improvement in reconstructed performance over BCS and 3D-CSC methods on most test datasets. It is worth noting that the SSIM of DLTG is higher in some cases than DD-UGM. The reason for this phenomenon is that DLTG algorithm conducts accurate imaging reconstruction through multiple complex parameter settings. Hence, it also has short comings of poor generalization ability and weak robustness. In contrast, the proposed method DD-UGM can be applied to a variety of imaging scenarios with the flexibility of the universal generative model without retraining the model. At 10-fold acceleration, it can be clearly seen that our proposed DD-UGM still achieves competitive quantitative performance. It follows then that the proposed method can effectively utilize learned dual-domain complementary prior information and low-rank information to improve the reconstruction quality consequently.

Figs. 6-7 demonstrate the reconstructed data from 8-fold and 10-fold acceleration, along with corresponding error maps for each method. Furthermore, the y-t image (extraction of the 100th slice along the y and temporal dimensions) is also given for each signal to show the reconstruction performance in the temporal dimension. From the error map and y-t view, we observe that the proposed method still provides comparable reconstruction results even at higher accelerations. Especially on the region of interest and around the edge of the papillary muscle, the proposed approach reduces most of the errors, while there are relatively higher residual errors in other comparison methods. Generally, DD-UGM can reduce data redundancy and learn the low-rank structure of the signal, which is regarded as a continuation of features and exhibits better performance in reconstruction.

TABLE I
THE AVERAGE PSNR, SSIM, AND MSE ($*e^{-4}$) OF COMPARISON METHOD AND THE PROPOSED METHOD ON THE SINGLE-COIL CARDIAC CINE TEST DATASET WITH DIFFERENT ACCELERATIONS (8-FOLD AND 10-FOLD) AT CARTESIAN SAMPLING PATTERN.

|  | Method | 201 | 202 | 203 | 204 | 205 | 206 | 207 | 208 | 209 | 210 | Average |
|---|---|---|---|---|---|---|---|---|---|---|---|---|
| **Cartesian R = 8** | BCS | 27.39 | 27.11 | 26.98 | 26.77 | 26.84 | 27.25 | 26.84 | 26.83 | 27.46 | 26.87 | 27.03 |
|  |  | 0.7513 | 0.7323 | 0.7002 | 0.6909 | 0.6756 | 0.6795 | 0.6552 | 0.6846 | 0.7117 | 0.6820 | 0.6963 |
|  |  | 18.394 | 19.521 | 20.819 | 21.900 | 21.150 | 18.943 | 20.884 | 21.204 | 18.169 | 20.844 | 20.183 |
|  | DLTG | 31.48 | 28.62 | 27.74 | 28.04 | 28.79 | 27.88 | 28.28 | 26.10 | 27.27 | 29.40 | 28.36 |
|  |  | 0.8489 | 0.8061 | **0.7933** | **0.7892** | 0.7732 | 0.7714 | **0.7539** | **0.7741** | **0.7966** | 0.7797 | 0.7886 |
|  |  | 7.152 | 13.829 | **17.449** | 18.863 | 14.434 | 17.811 | 15.150 | 26.262 | 19.552 | 11.825 | 16.233 |
|  | k-t SLR | 31.14 | 29.51 | 25.85 | 27.84 | 28.74 | 28.06 | 28.03 | 26.85 | 27.41 | 29.45 | 28.29 |
|  |  | 0.8498 | 0.8183 | 0.7745 | 0.7830 | 0.7694 | 0.7693 | 0.7509 | 0.7654 | 0.7904 | 0.7784 | 0.7849 |
|  |  | 7.707 | 11.351 | 27.208 | 19.060 | 14.259 | 16.430 | 16.162 | 21.856 | 19.361 | 11.586 | 16.498 |
|  | 3D-CSC | 25.99 | 26.20 | 24.17 | 24.98 | 25.86 | 26.01 | 25.82 | 24.73 | 25.09 | 24.97 | 25.38 |
|  |  | 0.7129 | 0.6926 | 0.6659 | 0.6515 | 0.6558 | 0.6462 | 0.6346 | 0.6440 | 0.6831 | 0.6021 | 0.6589 |
|  |  | 25.262 | 24.338 | 39.359 | 33.248 | 26.565 | 25.436 | 26.546 | 34.868 | 32.009 | 32.256 | 29.989 |
|  | DD-UGM | **32.47** | **30.79** | **28.12** | **28.05** | **29.09** | **28.81** | **29.22** | **27.93** | **28.45** | **29.47** | **29.24** |
|  |  | **0.8733** | **0.8369** | 0.7829 | 0.7872 | **0.7792** | **0.7743** | 0.7529 | 0.7645 | 0.7928 | **0.7860** | **0.7930** |
|  |  | **5.693** | **8.368** | 17.823 | **16.991** | **12.931** | **13.923** | **12.112** | **17.259** | **15.689** | **11.475** | **13.226** |
| **Cartesian R = 10** | BCS | 27.02 | 26.53 | 26.36 | 26.06 | 26.32 | 26.43 | 26.31 | 26.44 | 26.68 | 26.35 | 26.45 |
|  |  | 0.7290 | 0.7115 | 0.6805 | 0.6691 | 0.6530 | 0.6508 | 0.6350 | 0.6656 | 0.6885 | 0.6592 | 0.6742 |
|  |  | 19.878 | 22.379 | 23.795 | 25.578 | 23.796 | 23.103 | 23.645 | 23.534 | 22.047 | 23.557 | 23.131 |
|  | DLTG | 30.04 | 27.66 | 27.12 | 27.33 | 28.46 | 27.43 | 27.46 | 26.20 | 27.01 | 28.99 | 27.77 |
|  |  | 0.8288 | 0.7813 | **0.7793** | **0.7775** | 0.7601 | 0.7547 | 0.7305 | **0.7618** | **0.7845** | 0.7628 | 0.7721 |
|  |  | 9.966 | 17.326 | **20.052** | 21.128 | 15.164 | 19.781 | 18.344 | 25.361 | 20.459 | 12.728 | 18.031 |
|  | k-t SLR | 29.58 | 28.68 | 25.53 | 27.38 | 28.27 | 27.38 | 27.34 | 26.29 | 27.03 | 28.66 | 27.61 |
|  |  | 0.8293 | 0.7923 | 0.7575 | 0.7638 | 0.7535 | 0.7485 | 0.7276 | 0.7446 | 0.7726 | 0.7560 | 0.7646 |
|  |  | 11.030 | 13.604 | 28.846 | 20.719 | 15.491 | 19.456 | 18.762 | 25.003 | 20.526 | 13.884 | 18.732 |
|  | 3D-CSC | 26.75 | 26.15 | 24.40 | 24.93 | 25.41 | 25.52 | 25.63 | 24.52 | 25.31 | 25.83 | 25.45 |
|  |  | 0.7234 | 0.6933 | 0.6628 | 0.6479 | 0.6333 | 0.6327 | 0.6298 | 0.6500 | 0.6758 | 0.6466 | 0.6596 |
|  |  | 21.237 | 24.624 | 37.809 | 33.679 | 29.794 | 28.346 | 27.630 | 36.523 | 30.402 | 26.663 | 29.671 |
|  | DD-UGM | **31.77** | **30.26** | **26.83** | **27.71** | **28.79** | **28.71** | **28.85** | **27.27** | **28.01** | **29.35** | **28.75** |
|  |  | **0.8611** | **0.8249** | 0.7605 | 0.7707 | **0.7649** | **0.7597** | **0.7368** | 0.7458 | 0.7776 | **0.7724** | **0.7774** |
|  |  | **6.676** | **9.470** | 22.272 | **17.847** | **13.493** | **13.968** | **13.284** | **19.811** | **16.401** | **11.732** | **14.495** |

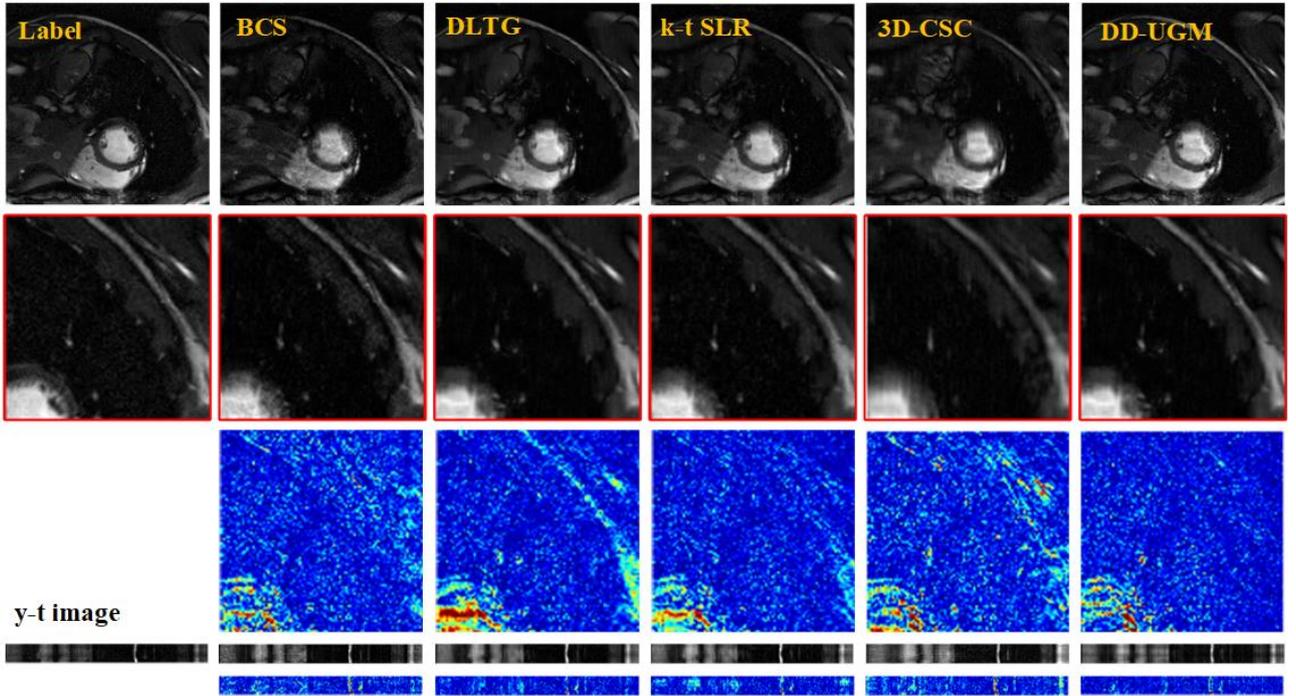

**Fig. 6.** The reconstruction results of the different methods (BCS, DLTG, k-t SLR, 3D-CSC, and the proposed DD-UGM) on the single-coil cardiac cine test 210 dataset with 8-fold acceleration and Cartesian sampling pattern. The red boxes indicate the region of interest and the intensity of residual maps is five times magnified.

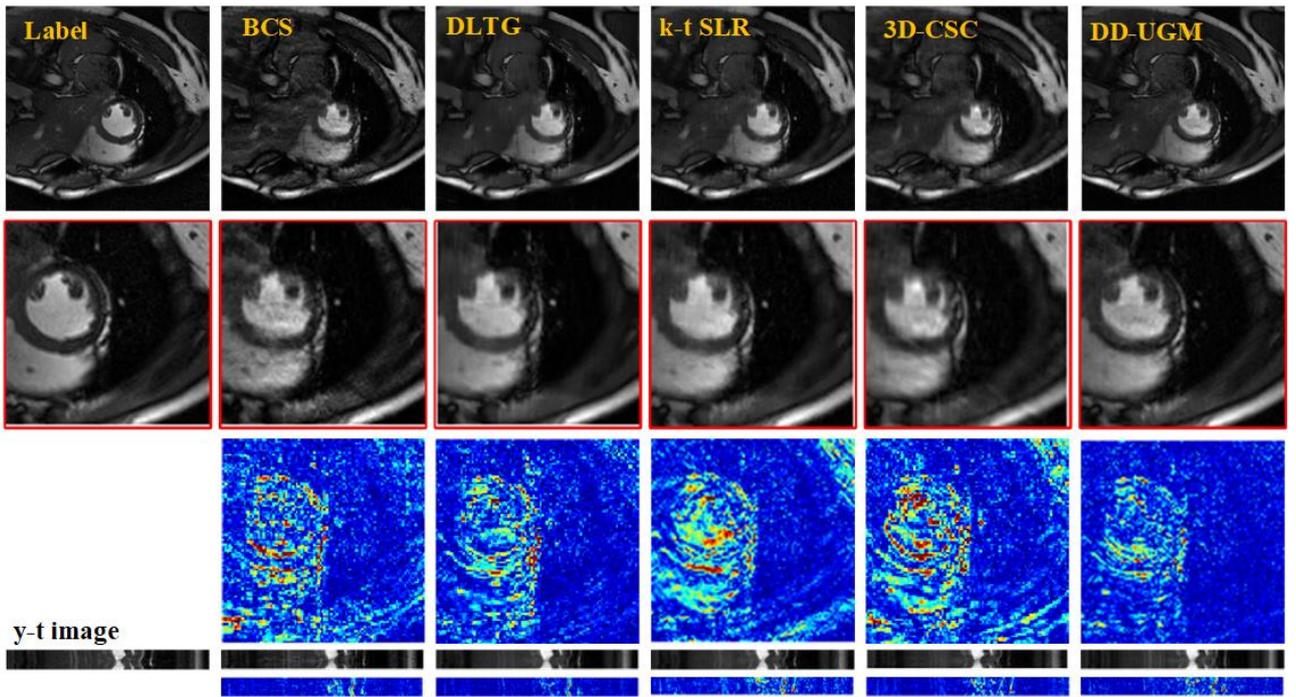

**Fig. 7.** The reconstruction results of the different methods (BCS, DLTG, k-t SLR, 3D-CSC, and the proposed DD-UGM) on the single-coil cardiac cine test 201 dataset with 10-fold acceleration and Cartesian sampling pattern. The red boxes indicate the region of interest and the intensity of residual maps is five times magnified.

## V. Discussion

### A. Comparison of Different Sampling Patterns

The models in the above sections are all trained with a 1D Cartesian sampling pattern. Although our proposed framework is based on a time-interleaved sampling scheme, network training, and testing can be conducted with any sampling patterns. In this section, we train the model under different sampling patterns in the single-channel scenario. For the Cartesian trajectory, we use equal spacing within each frame. The reconstruction results under Radial and Cartesian sampling at 10-fold and 8-fold acceleration factors are shown in Fig. 8. From the reconstructed images and their corresponding error maps of different reconstruction methods, DD-UGM achieves reasonable reconstruction results when using Radial and Cartesian sampling patterns. Furthermore, the quantitative indicators tabulated in Table II confirm that our proposed DD-UGM obtains better quantitative performance than k-t SLR under each under-sampling mask. Therefore, the proposed method trained on one sampling pattern can well be generalized to other sampling patterns.

TABLE II
THE AVERAGE PSNR, SSIM, AND MSE ($*E^{-4}$) OF THE PROPOSED METHOD AT $R = 10$ AND $R = 8$ ON THE SINGLE-COIL CARDIAC CINE TEST DATASET UNDER DIFFERENT SAMPLING PATTERNS.

|  | Method | 201 | 202 | 203 | 204 | 205 | 206 | 207 | 208 | 209 | 210 | Average |
|---|---|---|---|---|---|---|---|---|---|---|---|---|
| Radial $R = 10$ | k-t SLR | 31.23 0.8158 7.564 | 30.39 0.7936 9.153 | **29.77** **0.7550** **11.089** | 29.07 0.7562 12.917 | 29.50 0.7361 11.371 | 29.08 0.7336 13.092 | 28.09 0.7099 15.759 | **29.64** **0.7393** **11.103** | 29.20 0.7615 12.552 | 29.54 0.7463 11.309 | **29.55** 0.7547 11.590 |
|  | DD-UGM | **32.14** **0.8559** **6.132** | **30.72** **0.8206** **8.506** | 27.36 0.7688 19.247 | 27.87 0.7714 17.998 | 28.81 0.7611 13.667 | 28.21 0.7529 16.171 | **28.77** **0.7302** **13.493** | 26.84 0.7475 22.803 | 28.08 0.7811 16.697 | **29.61** **0.7698** **10.996** | 28.84 0.7759 14.571 |
| Cartesian $R = 8$ | k-t SLR | 31.14 0.8498 7.707 | 29.51 0.8183 11.351 | 25.85 0.7745 27.208 | 27.84 0.7830 19.060 | 28.74 0.7694 14.259 | 28.06 0.7693 16.430 | 28.03 0.7509 16.162 | 26.85 0.7654 21.856 | 27.41 0.7904 19.361 | 29.45 0.7784 11.586 | 28.29 0.7849 16.498 |
|  | DD-UGM | **32.47** **0.8733** **5.693** | **30.79** **0.8369** **8.368** | **28.12** **0.7829** **17.823** | **28.05** **0.7872** **16.991** | **29.09** **0.7792** **12.931** | **28.81** **0.7743** **13.923** | **29.22** **0.7529** **12.112** | **27.93** **0.7645** **17.259** | **28.45** **0.7928** **15.689** | **29.47** **0.7860** **11.475** | **29.24** **0.7930** **13.226** |

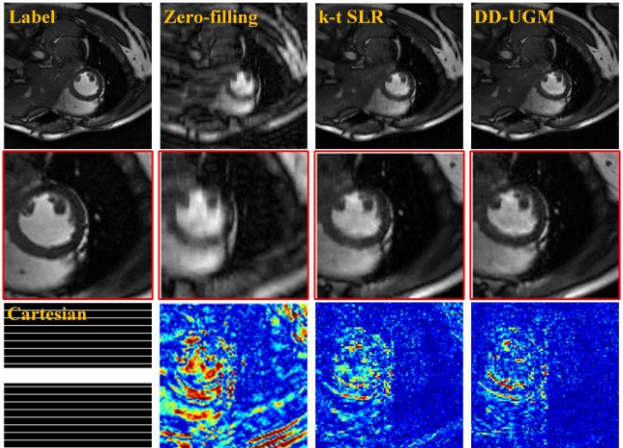

**Fig. 8.** Reconstruction results of k-t SLR and DD-UGM in the single-coil scenario under Cartesian mask at 8-fold. The red boxes indicate the region of interest and the intensity of residual maps is five times magnified.

### B. Comparison between Single-domain and Dual-domain

In this subsection, the effect of reconstruction in single-domain and dual-domain is discussed. Before showing the experimental results, we first discuss the essential difference between dual-domain and single-domain. On the one hand, the dual-domain needs to train k-space and image domain models while the single-domain only trains the model through a kind of data. On the other hand, the single-domain only requires a sort of data for iterative reconstruction while the dual-domain needs to input k-space and image data into the corresponding network at the same time. Hence, the reconstructed results of the different domains are weighted after each iteration.

The comparison of the visualization results of the different domains at 8-fold acceleration is shown in Fig. 9. The first column is the actual dynamic MR image. From left to right are the reconstruction results of zero filling, single-domain, and dual-domain respectively. Fig. 9 clearly reveals that the dual-domain model can offer reconstructions that are less blurred and have fewer alias artifacts when compared to the single-domain.

Besides, the quantitative measurements can be found in Table III. One can see that the dual-domain model achieves optimal quantitative evaluations over the single-domain. Accordingly, the dual-domain is superior in both visual results and quantitative indicators. These indicators indicate the dual-domain model could effectively learn complementary information in different models and low-rank prior knowledge to improve DMRI reconstruction. Therefore, we choose dual-domain model to reconstruct the image.

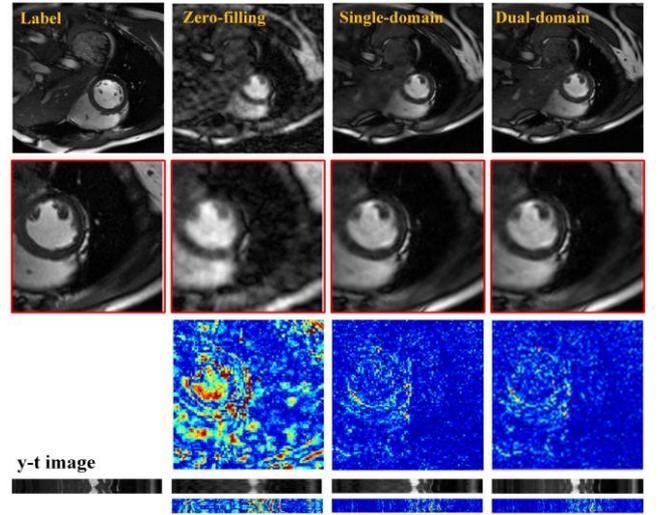

**Fig. 9.** The reconstruction results of the different domains (Zero-filling, Single-domain, Dual-domain) on the single-coil cardiac cine test 201 dataset with 8-fold acceleration at Radial sampling pattern.

TABLE III
THE AVERAGE PSNR, SSIM, AND MSE ($*E^{-4}$) OF THE PROPOSED METHOD AT $R = 8$ AND $R = 10$ ON THE SINGLE-COIL CARDIAC CINE TEST DATASET UNDER RADIAL SAMPLING PATTERNS.

|  | Method | 201 | 202 | 203 | 204 | 205 | 206 | 207 | 208 | 209 | 210 | Average |
|---|---|---|---|---|---|---|---|---|---|---|---|---|
| Radial $R = 8$ | Single-domain | 32.44 0.8592 5.739 | 31.14 0.8285 7.711 | 27.40 0.7645 19.977 | 27.90 0.7738 18.099 | 28.95 0.7708 13.239 | 28.66 0.7596 14.996 | **29.22** **0.7383** **12.108** | 27.35 0.7465 20.011 | 28.08 0.7805 16.709 | 29.80 0.7782 10.606 | 29.0942 0.7800 13.9193 |
|  | Dual-domain | **32.54** **0.8611** **5.609** | **31.16** **0.8303** **7.685** | **27.57** **0.7763** **19.305** | **28.00** **0.7806** **17.818** | **29.01** **0.7695** **12.941** | **28.70** **0.7613** **14.869** | 29.17 0.7362 12.243 | **27.45** **0.7592** **19.670** | **28.20** **0.7893** **16.383** | **29.81** **0.7786** **10.591** | **29.1600** **0.7842** **13.7114** |
| Cartesian $R = 10$ | Single-domain | 31.69 0.8592 6.798 | 30.24 0.8220 9.555 | 26.82 0.7461 22.261 | 27.71 **0.7734** 18.192 | 28.72 0.7595 13.717 | 28.61 0.7512 14.298 | 28.85 0.7366 13.278 | 27.21 0.7332 20.025 | 27.97 0.7652 16.502 | 29.30 0.7705 11.889 | 28.7103 0.7717 14.6514 |
|  | Dual-domain | **31.77** **0.8611** **6.676** | **30.26** **0.8249** **9.470** | **26.83** **0.7605** **22.272** | **27.72** 0.7707 **17.847** | **28.79** **0.7649** **13.493** | **28.71** **0.7597** **13.968** | **28.86** **0.7368** **13.284** | **27.27** **0.7458** **19.811** | **28.01** **0.7776** **16.401** | **29.35** **0.7724** **11.732** | **28.75** **0.7774** **14.4954** |

## C. Convergence Analysis and Computational Cost

In this section, we experimentally investigate the convergence of DD-UGM with the number of iterations. Taking the 1st sequence on test dataset 201 as an example, Fig. 10 depicts the comparative plot of convergence tendency of PSNR curves to the number of iterative steps, reflecting the stabilities of reconstructing. As seen in Fig. 7, DD-UGM converges rapidly to near the peak value on account of the existence of error threshold values. Hence, it can be seen that the proposed DD-UGM greatly shortens the reconstruction time and is more stable because of its robustness to noise, which makes a faster reconstruction possible.

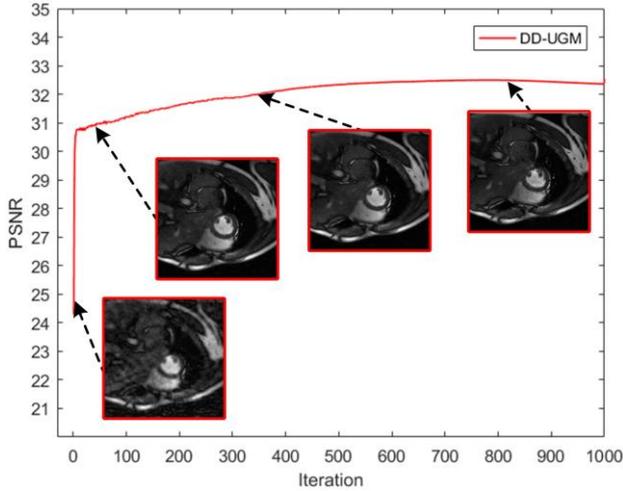

**Fig. 10.** Convergence curve of single-domain and dual-domain in terms of PSNR versus iterations.

## D. Reconstruction on Real-Time OCMR

To prove the universality of the model, we also apply DD-UGM to reconstruct the free-breathing full-sampled cardiac dataset OCMR [46]. We compare DD-UGM with a state-of-the-art method L+S [11] in different sampling patterns. The reconstruction results of these methods at 8-fold acceleration factor are shown in Fig. 11. Notably, the spatial resolution is lower than that of the data in the retrospective study, so the image quality in this experiment is not as good as in the former images. The qualitative results demonstrate that DD-UGM can still successfully utilize the complementary dual-domain prior information and low-rank prior of dynamic signals to improve the reconstruction results than L+S method under four sampling patterns. Specifically, DD-UGM has fewer streaking artifacts and less spatial blurring.

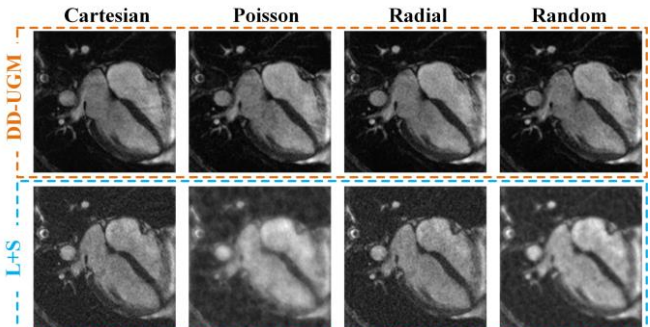

**Fig. 11.** The reconstruction results of the proposed DD-UGM and L+S under different sampling patterns at $R = 8$ on full-sampled cardiac dataset.

## VI. CONCLUSIONS

This work presented a dual-domain generative model for DMRI reconstruction with score-based prior embedded. In detail, we proposed to restore both k-space and image domains recurrently through score-based model with VE-SDE. Especially, it was better to generate samples and minimizes the data consistency cost via VE-SDE. Due to the uniqueness of the data in the k-space domain, the weighted technology was employed to make the output of high-frequency information sparser and facilitate the training of k-space domain model. Moreover, the low-rank information prior through the SVD algorithm was embedded at the reconstruction phase of k-space domain to deeply guide restorations from highly under-sampled measurements. Overall, our approach is universal and flexible, which can reconstruct images of any number of frames with only one frame of training. Extensive experimental results also demonstrated that our method could improve the reconstruction results both qualitatively and quantitatively. Future work includes extending the proposed framework to dynamic PET image reconstruction and further improving the redundancies in the multi-channel data for further improvements. Moreover, the connection of DMRI and dynamic PET images for joint reconstruction will also be the subject of further research.